# CDW Ordering in Stripe Phase of Underdoped Cuprates

A. Mourachkine

*Université Libre de Bruxelles, Service de Physique des Solides, CP233, Boulevard du Triomphe, B-1050 Brussels, Belgium*

**Abstract**

The in-plane resistivity $\rho_{ab}(T)$ and out-of-plane resistivity $\rho_c(T)$ of non-superconducting $R$Ba$_2$Cu$_3$O$_{6+x}$ ($R$BCO) ($R$ = Y, Tm) and Fe-doped Bi$_2$Sr$_2$CaCu$_2$O$_{8+x}$ (Bi2212) single crystals are discussed. The comparison of electrical transport properties of the cuprates and quasi-one dimensional (1D) (TMTSF)$_2$PF$_6$ organic conductor suggests that $R$BCO and Bi2212 exhibit 1D transport properties, and the step rise at low temperatures in the $\rho_{ab}$ and $\rho_c$ of the cuprates and quasi-1D organic conductor is due to charge-density-wave ordering. We discuss also phonon-electron interactions in cuprates at low temperatures.

*Keywords*: A. Superconductors, 2D cuprate SCs, 1D organic conductors, Electrical properties

## 1. Introduction

There is no consensus on the mechanism of superconductivity (SC) in copper-oxides. There is a clear microscopic difference between the conventional SC and high-$T_c$ SC, namely that they have different origins and that different criteria are required for the high-$T_c$ SC than for the classical SC. The key-stone in the understanding of the high-$T_c$ SC mechanism is the understanding of *the normal state* of cuprates, which is also unique [1]. There are two ways to understand the properties of cuprates: to study them directly *or* to compare them with physical properties of systems which we understand much better.

There is clear evidence for charge stripe [2] formation in SC La$_{1.6-x}$Nd$_{0.4}$Sr$_x$CuO$_4$ (Nd-doped LSCO) [3-6] and YBCO [7-10]. There are also strong indications that stripes exist in SC Bi2212 [11,12]. Holes induced in CuO$_2$ planes segregate into periodically-spaced stripes that separate antiferromagnetic (AF) isolating domains. Recent magnetoresistance measurements in *non*-SC YBCO show the presence of charge stripes which can be directed with an external magnetic field [13]. 1D electronic systems are expected to show unique transport behavior as a consequence of the Coulomb interaction between carriers [14]. Some of them exhibit peculiar electronic behavior such as SC or the Peierls transition. The

Peierls transition [14] is associated with a charge density wave (CDW) which is a hybrid of electron wave and phonon. Theoretically, the low temperature transport properties of interacting 1D-electron systems are described in term of a Luttinger liquid rather than an usual Fermi liquid [15]. It is interesting to note that Luttinger liquids do not have dominant CDW correlations, however, a spin-gapped Luttinger liquid does [16].

Thus, if measurements in cuprates show the presence of quasi-1D charge stripes [3-13], it is absolutely logic to compare transport properties of cuprates with similar properties of quasi-1D metals. Most of the non-CDW quasi-1D metals are organic. In the present work, we compare temperature dependencies of the in-plane resistivity $\rho_{ab}(T)$ and out-of-plane resistivity $\rho_c(T)$ of non-SC $R$BCO ($R$ = Y, Tm) and Fe-doped Bi2212 single crystals and a Bechgaard salt $(TMTSF)_2PF_6$. The comparison suggests that $R$BCO and Bi2212 exhibit 1D transport properties, and the step rise at low temperatures in the $\rho_{ab}$ and $\rho_c$ of the cuprates and $(TMTSF)_2PF_6$ is due to a CDW instability. We discuss also phonon-electron interactions in cuprates.

## 2. Bechgaard salt $(TMTSF)_2PF_6$

Before analyzing transport properties of cuprates let us discuss first the properties of quasi-1D $(TMTSF)_2PF_6$ organic conductor, where TMTSF and $PF_6$ are tetramethyltetraselenafulvalene and hexafluorophosphate, respectively.

Recently, Bechgaard salts have attracted much attention due to their unique conduction mechanism. Bechgaard salts are made of packed and flat charged molecules in an organic salt [17-19]. Since these flat molecules interact weakly in the solid state, they retain their flat shape, which results in a good intermolecular overlap of the electron clouds along the packing direction, making electron delocalisation possible along that direction. The Bechgaard salt $(TMTSF)_2PF_6$ is one of the most studied organic conductors. These molecules naturally stack in 1D structure with two TMTSF molecules and one $(PF_6)^-$ ion per unit cell. The overlap between the electron clouds of neighboring molecules in the stacking *a* direction is about 500 times larger than that along the transverse *c* direction (along the long axis of the individual molecules) [17-19]. At high pressure, below 10 K, the Bechgaard salt becomes SC. The electrical transport properties of the Bechgaard salt along the stacks and perpendicular to the stacks are an excellent starting point for studying any system which is expected to have 1D physical properties since the 1D structure of the Bechgaard salt is obvious.

Figure 1 shows temperature dependencies of two resistivities of the Bechgaard salt in two different directions [19]. In Fig. 1, the longitudinal resistivity, $\rho_a$, along the stacks exhibits metallic behavior by falling with temperature almost as $T^2$. The transverse resistivity, $\rho_c$, increases first as $T$ gets smaller. This is *most likely* an indication of Luttinger-liquid behavior [18,19]. However, below 90 K, the $\rho_c$ falls with temperature signaling the onset of instability in the Luttinger liquid and the progressive recovery of a Fermi liquid [18,19]. The steep rise at low temperatures is due to the transition from a metal to an insulator [18,19]. This transition can occur, in general, due to either CDW *or* spin density wave (SDW) instability. It is important to note that the transition to the insulating state in the $\rho_a$ and $\rho_c$ occurs at different temperatures: $T\rho_a < T\rho_c$. The transverse resistivity of $(TMTSF)_2PF_6$ in the *b* direction, $\rho_b(T)$, measured by a microwave technique shows also the existence of the maximum around 40 K which is somewhat lower than 90 K shown by $\rho_c(T)$ in Fig. 1 [17].

## 3. Heavily underdoped *R*BCO

We discuss now the electrical transport properties of heavily underdoped non-SC *R*BCO. Figure 2(a) shows temperature dependencies of the in-plane, $\rho_{ab}$, and out-of-plane, $\rho_c$, resistivities in a non-SC $TmBa_2Cu_3O_{6.37}$ (TmBCO) single crystal [20]. By comparing Figs. 1 and 2(a) one can find an obvious similarity between the temperature dependencies of the two resistivities. In Fig. 2(a), between 125 K and 327 K, the $\rho_c$ exhibits a temerature dependence similar to the behavior the $\rho_c$ of the Bechgaard salt (Luttinger-liquid behavior). The $\rho_{ab}(T)$ is proportional to $T^{1.8}$ for $T$ < 327 K and to $T^2$ for $T$  327 K. The $\rho_{ab}$ does not show the rise at low temperature because, most likely, the transition temperature to the insulating state is lower than the minimum measured temperature. This transition to the insulating phase can be observed in the $\rho_{ab}$ of heavily underdoped YBCO single crystals, which are shown in Fig. 2(b) [13]. Both the YBCO crystals with *x* = 0.30 and *x* = 0.32 are AF below 200 K. Consequently, the insulating state at low temperature in Fig. 2(b) can be *only* explained by the appearance of a CDW order since the SDW order already exists in the YBCO samples below 200 K. By comparing the temperature dependencies of the $\rho_a$ shown in Fig. 1 and $\rho_{ab}$ in Fig. 2(b) one can find again an obvious similarity. Thus, the electrical transport in *R*BCO exhibits 1D character. In Fig. 2(a), one can mentioned that there is a transition in both the $\rho_{ab}$ and $\rho_c$ at $T^* = 327$ K. It is most likely that $T^*$ corresponds to the formation of charge stripes in TmBCO.

## 4. Fe-doped Bi2212

Here we discuss briefly the out-of-plane resistivity, $\rho_c(T)$, of a Fe-doped Bi2212 single crystal. Figure 3 shows the temperature dependence of the $\rho_c$. Measurements have been performed by four-probe method. The content of Fe with respect to Cu is a few %. The Fe-Bi2212 single crystals were fabricated by self-flux method to perform tunneling measurements [21,22]. However, the crystals happen to be not SC. In Fig. 3, the temperature dependence of the $\rho_c$ is very peculiar. Between 32 K and 290 K, the temperature dependence of the $\rho_c$ is *typical* for underdoped Bi2212 single crystals [23,24]. At 32 K, the Fe-doped Bi2212 sample becomes SC in terms of the limits of the voltage resolution in the experiment (a few nV). Below 31.5 K, suddenly, the sample becomes insulating [25]. In fact, it is also indication of 1D physics since, in 1D physics, a SC and CDW instabilities are treated on equal footing, thus, a SC instability can be easily transformed into a CDW instability, and vice versa [16,26].

We attempt to explain the temperature dependence of the $\rho_c$ shown in Fig. 3. With decrease of temperature from 290 K to 45 K, charge carriers are more confined into $CuO_2$ planes where they gather into 1D stripes [13]. Between 45 K and 32 K, the decrease of the $\rho_c$ occurs due to either the progressive recovery of a Fermi liquid [18,19] or the occurrence of SC fluctuations. At 32 K when the sample become SC or almost SC, stripes are redistributed in such way that all of them become pinned by Fe impurities located in $CuO_2$ planes [3]. The latter immediately leads to the appearance of a CDW order [26].

## 5. Discussion

Let us consider a few other facts. The comparison of temperature dependencies of the in-plane resistivity in a $YBa_2Cu_3O_8$ single crystal and resistivity along the spin chains in a even-chain spin-ladder $Sr_{2.5}Ca_{11.5}Cu_{34}O_{41}$ single crystal, both scaled to the temperature at which the pseudogap [1] appears, shows that the temperature depndencies of the resistivities are identical [10]. Spin-ladders have obvious 1D structure. The latter fact points out that the in-plane electrical transport in $YBa_2Cu_3O_8$ has not 2D but 1D character. Thus, it seems that the presence of charge stripes is intrinsic to all cuprates. It is interesting to note that the carrier densities along stripes in cuprates and along stacks in

TMTSF-DMTCNQ organic SC are the same: 0.5 hole/Cu in cuprates [3] and 0.5 hole/molecule in TMTSF-DMTCNQ organic SC [17].

In Section 3, we found that the step rise in the in-plane resistivities of the YBCO single crystals at low temperature can be only interpreted by the appearance of a CDW order. The appearance of CDW in cuprates at low temperature indicates that phonon-electron interactions are strong at low temperatures comparing with other energy scales. Measurements of the nuclear quadrupole resonance in YBCO with $T_c$ = 89-90 K reveal unusual features below 35 K, which were attributed to a CDW ordering [27]. Tunneling measurements in LSCO show that phonons couple with electrons below $T_c$ = 34.4 K [28]. Thus, it seems that phonon-electron interactions are important in cuprates and have to be taken into account at low temperatures (< 35 K).

## 6. Conclusions

We discussed the in-plane resistivity $\rho_{ab}(T)$ and out-of-plane resistivity $\rho_c(T)$ of non-SC $R$BCO ($R$ = Y, Tm) and Fe-doped Bi2212 single crystals. The comparison of electrical transport properties of the cuprates and quasi-1D (TMTSF)$_2$PF$_6$ organic conductor suggests that $R$BCO and Bi2212 exhibit 1D transport properties. The step rise at low temperatures in the $\rho_{ab}$ and $\rho_c$ of the cuprates and quasi-1D organic conductor is due to the CDW ordering. We conclude that phonon-electron interactions in cuprates at low temperatures (< 35 K) are strong and have to be taken into account.

### Acknowledgments

I thank A. N. Lavrov for providing the original data shown in Fig. 2 and sending Ref. 13 prior to publication, and R. Deltour for a discussion. This work is supported by PAI 4/10.

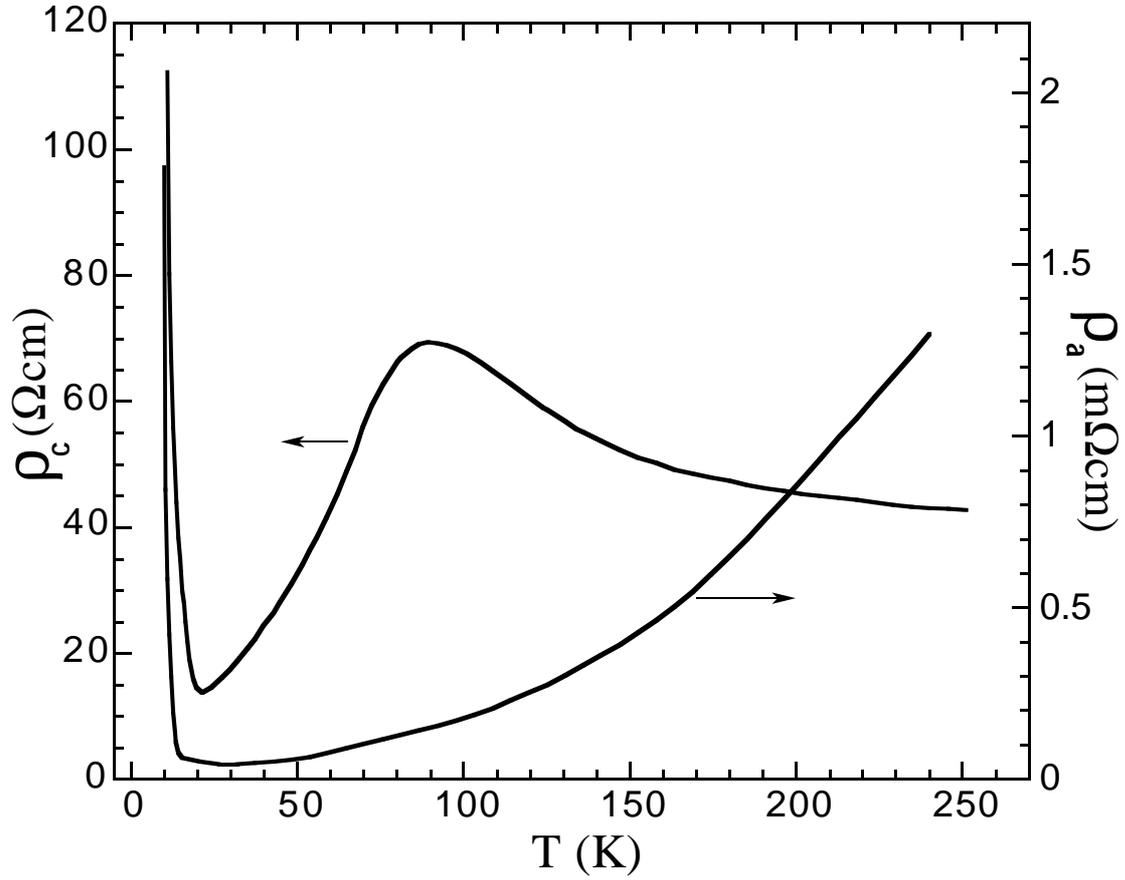

FIG. 1. Longitudinal, $\rho_a$, and transverse, $\rho_c$, resistivities as a function of temperature in $(TMTSF)_2PF_6$ (from Ref. 19).

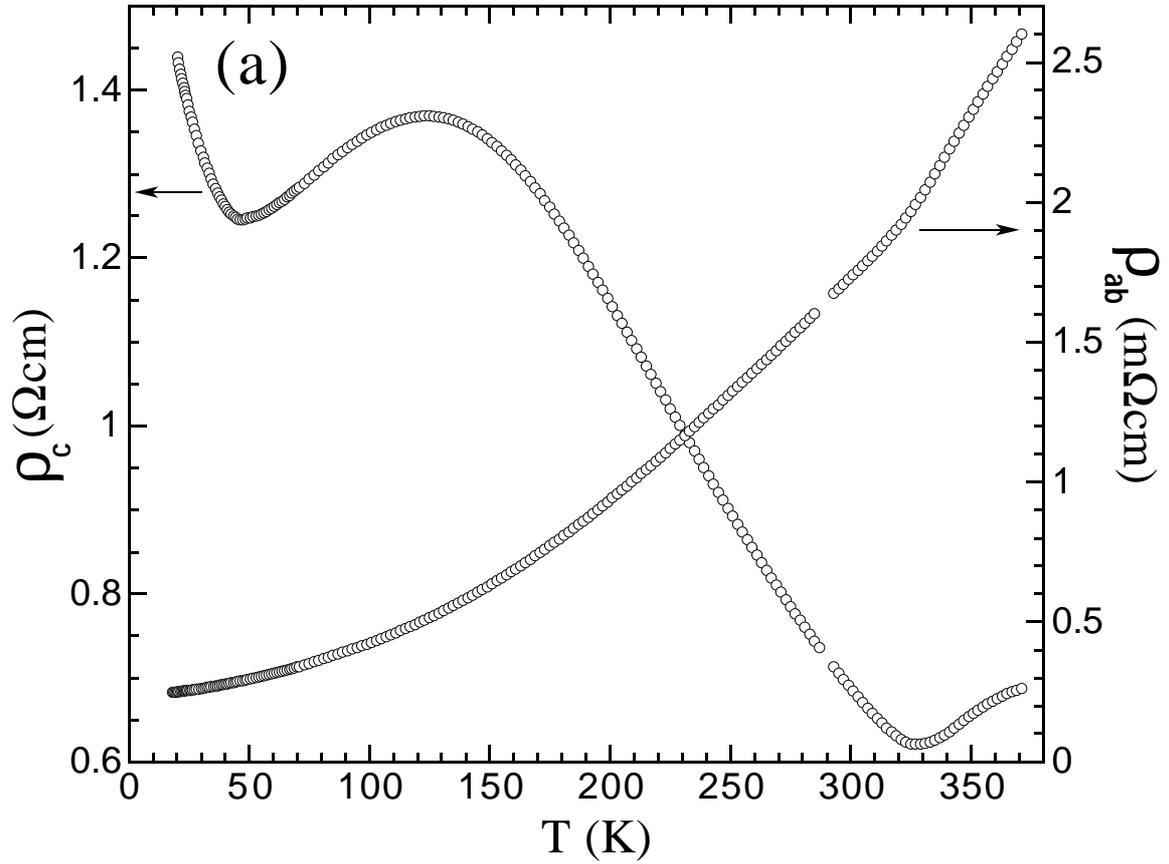

FIG. 2. (a) In-plane, $\rho_{ab}(T)$, and out-of-plane, $\rho_c(T)$, resistivities of a TmBCO single crystal (from Ref. 20);

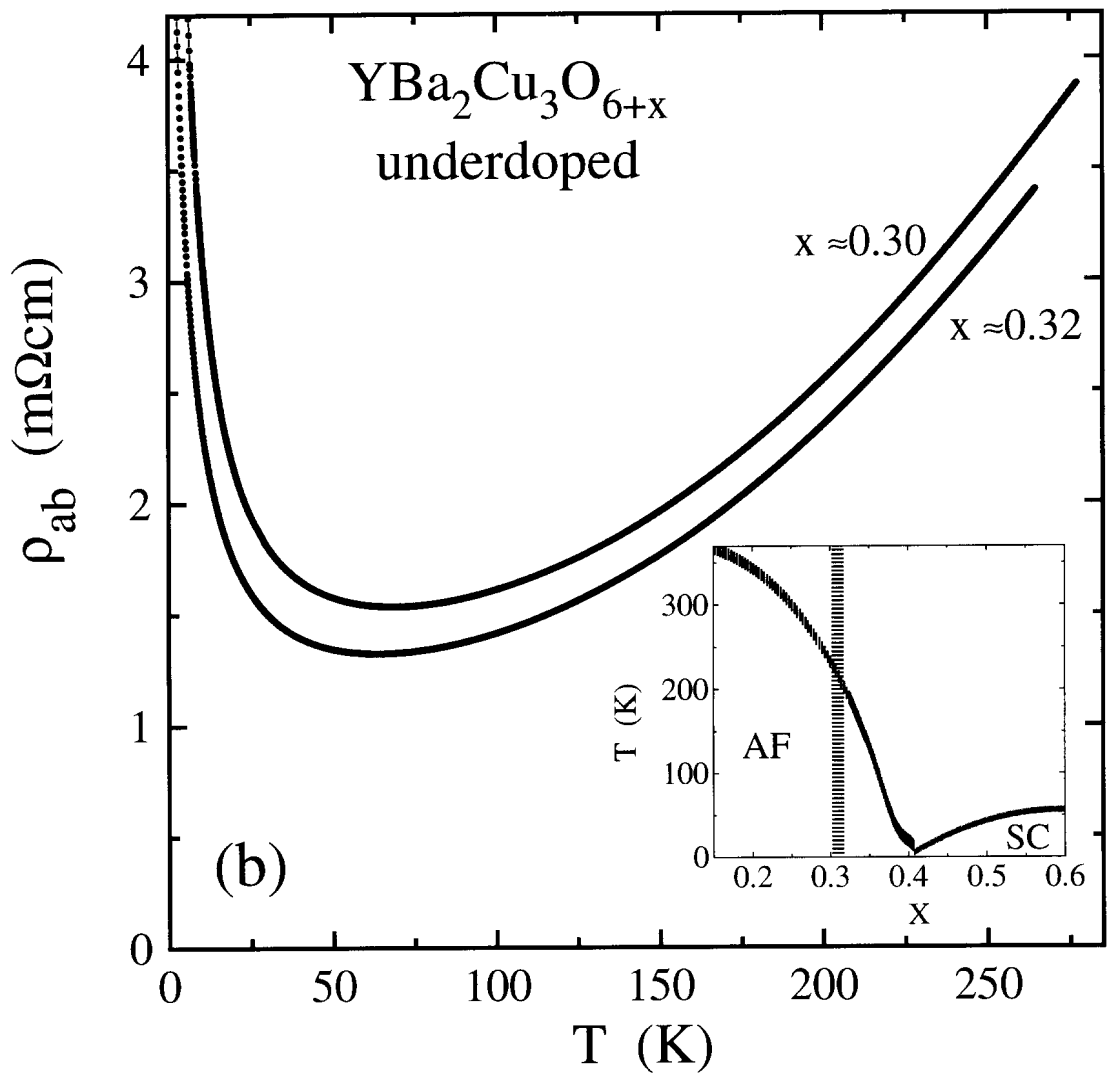

FIG. 2. (b) the $\rho_{ab}(T)$ of two YBCO single crystals with x = 0.3 and 0.32. Inset in the plot (b) illustrates the part of the phase diagram being studied (from Ref. 13).

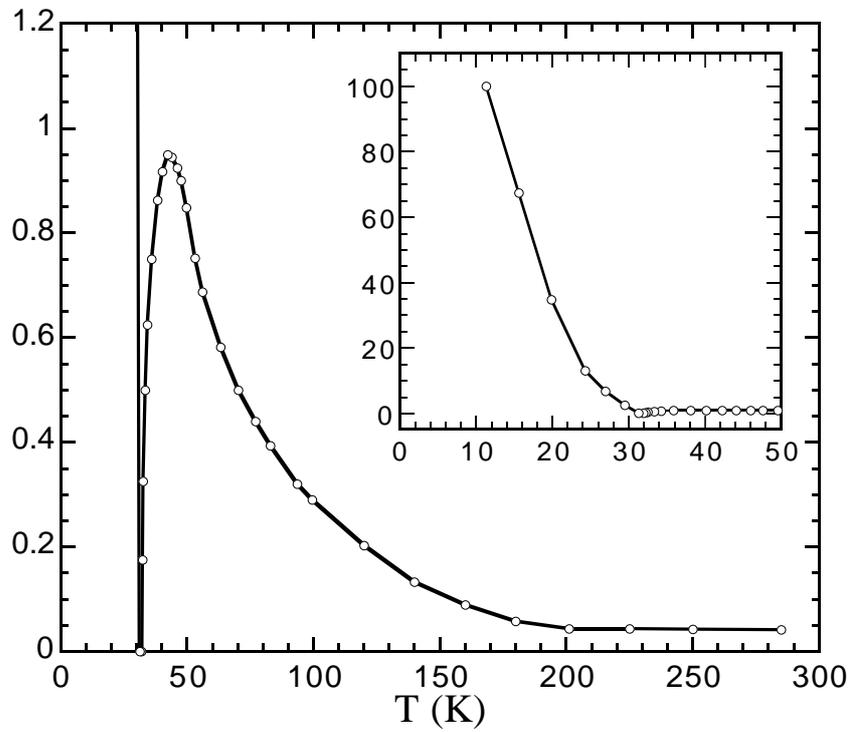

Figure 3 Out-of-plane, $\rho_c(T)$, resistivity of a Fe-doped Bi2212 single crystal. Inset: the $\rho_c(T)$ at low temperature (same axis parameters as main plot).